\makeatletter
\def\cl@chapter{}
\makeatother
\RequirePackage{fix-cm}

\documentclass[smallextended,numbook]{svjour3}

\usepackage{hyperref}
\usepackage{cleveref}

\usepackage{graphicx}
\graphicspath{{./assets/}}

\usepackage{textcomp}
\usepackage{xcolor}

\usepackage{cite}
\usepackage{url}

\usepackage[T1]{fontenc}

\definecolor{codegreen}{rgb}{0,0.6,0}
\definecolor{codegray}{rgb}{0.5,0.5,0.5}
\definecolor{codepurple}{rgb}{0.58,0,0.82}
\definecolor{backcolour}{rgb}{0.95,0.95,0.92}

\usepackage{listings}
\lstdefinestyle{mystyle}{
    backgroundcolor=\color{backcolour},
    commentstyle=\color{codegreen},
    keywordstyle=\color{magenta},
    numberstyle=\tiny\color{codegray},
    stringstyle=\color{codepurple},
    basicstyle=\ttfamily\footnotesize,
    breakatwhitespace=false,
    breaklines=true,
    captionpos=b,
    keepspaces=true,
    numbers=left,
    numbersep=5pt,
    showspaces=false,
    showstringspaces=false,
    showtabs=false,
    tabsize=2
}

\usepackage{xspace}

\usepackage{csquotes}
\usepackage{booktabs}
\usepackage{subcaption}
\usepackage{graphicx}
\usepackage[normalem]{ulem}

\definecolor{wzcolor}{RGB}{255,140,0}

\definecolor{mikecolor}{RGB}{255,140,0}

\newcommand{\rheader}[1]{{\noindent\emph{#1}: }}

\newcommand{\wzVTopUsersWithDotfilesStrict}{112\xspace}
\newcommand{\wzVTopUsersWithDotfiles}{129\xspace}
\newcommand{\wzVTotalDotfilesRepos}{167,663\xspace}
\newcommand{\wzVTotalClonedDotfilesRepos}{147,548\xspace}

\newcommand{\wzVDatasetDotfilesRepos}{3,305\xspace}

\newcommand{\wzVTSCount}{12,502\xspace}
\newcommand{\wzVTSUniqueCount}{3090\xspace}

\newcommand{\etal}{et~al.} % need manual '~' after etal

\newcommand{\dotrepos}{\emph{dotfiles} repositories\xspace}

\newcommand{\dotrepo}{\emph{dotfiles} repository\xspace}

\newcommand{\dotdataset}{\emph{dotfiles dataset}\xspace}
\newcommand{\dfs}{\emph{dotfiles}\xspace}
\newcommand{\Dfs}{\emph{Dotfiles}\xspace}
\newcommand{\df}{\emph{dotfile}\xspace}

\newcommand{\nix}{\emph{*NIX}\xspace}
\newcommand{\kshape}{\emph{K-Shape}\xspace}
\newcommand{\readme}{\emph{README}\xspace}
\newcommand{\github}{\emph{GitHub}\xspace}

\newcommand{\git}{\emph{Git}\xspace}
\newcommand{\emacs}{\emph{Emacs}\xspace}
\newcommand{\vim}{\emph{Vim}\xspace}
\newcommand{\vscode}{\emph{VSCode}\xspace}

\newcounter{observationNum}
\newcommand{\observe}[1]{
    \stepcounter{observationNum}
    \textbf{Observation \theobservationNum: #1}}

\newcounter{wzobs}[section]

\newcounter{challengeNum}
\newcommand{\challenge}[1]{
    \stepcounter{challengeNum}
    \textbf{Challenge \thechallengeNum: #1}}

\newcounter{opportunityNum}
\newcommand{\opportunity}[1]{
    \stepcounter{opportunityNum}
    \textbf{Opportunity \theopportunityNum: #1}}

\begin{document}

\title{An Empirical Study of Dotfiles Repositories Containing User-Specific Configuration Files}

\author{Wenhan Zhu \and
  Michael W. Godfrey
}

\institute{W. Zhu \at
  David R. Cheriton School of Computer Science \\
  \email{w65zhu@uwaterloo.ca}
  \and
  M. W. Godfrey \at
  David R. Cheriton School of Computer Science \\
  \email{migod@uwaterloo.ca}
}

\maketitle

\begin{abstract}

Storing user-specific configuration files in a ``dotfiles'' repository is a common practice among software developers, with hundreds of thousands choosing to publicly host their repositories on \github. 
This practice not only provides developers with a simple backup mechanism for their essential configuration files, but also facilitates sharing ideas and learning from others on how best to configure applications that are key to their daily workflows. 
However, our current understanding of these repository sharing practices is limited and mostly anecdotal. 
To address this gap, we conducted a study to delve deeper into this phenomenon. 
Beginning with collecting and analyzing publicly-hosted \dfs repositories on \github, we discovered that maintaining \dfs is widespread among developers. 
Notably, we found that 25.8\% of the top 500 most-starred \github users maintain some form of publicly accessible \dfs repository. 
Among these, configurations for text editors like \vim and shells such as \emph{bash} and \emph{zsh} are the most commonly tracked. 
Our analysis reveals that updating \dfs is primarily driven by the need to adjust configurations (63.3\%) and project meta-management (25.4\%). 
Surprisingly, we found no significant difference in the types of \dfs observed across code churn history patterns, suggesting that the frequency of \df modifications depends more on the developer than the properties of the specific \df and its associated application. 
Finally, we discuss the challenges associated with managing \dfs, including the necessity for a reliable and effective deployment mechanism, and how the insights gleaned from \dfs can inform tool designers by offering real-world usage information.

\end{abstract}

%-- Storing user-specific configuration files in a ``dotfiles'' repository is a common practice among software developers.
%-- Hundreds of thousands of developers have chosen to host their own \dotrepo publicly on \github.
%-- Storing these configuration files allows developers to have a backup of their important configurations, and sharing them publicly enables developers to learn from others.
%-- 
%-- Currently, we have only a limited and anecdotal understanding of the practices of storing and sharing user-specific configuration files.
%-- In this study, we aim to gain a better understanding of this phenomenon.
%-- We start by collecting and examining a set of publicly hosted \dotrepos from \github.
%-- We find that maintaining \dfs is a common practice among developers.
%-- 25.8\% of the top 500 most-starred users on \github own a variant of a \dotrepo.
%-- Configurations for text editors (e.g., \vim) and shells (e.g., \emph{bash}, \emph{zsh}) are the most commonly tracked \dfs.
%-- Adjusting configurations (63.3\%) and \dfs project-meta management (25.4\%) are the most common reasons for updating the \dotrepo.
%-- We also found no significant difference in the type of \dfs observed across the code churn history patterns, suggesting that how often are \dfs modified depends more on the developer than properties of the specific \df.
%-- Finally, we discuss the challenges in managing \dfs, such as the need for a reliable and effective deployment mechanism, and how the information contained in \dfs can benefit tool designers by providing real-world usage information.


\section{Introduction}

% dotfiles are important and we should care about them
Tools are essential in software development, and studies have shown that familiarity with tools can significantly increase developer productivity~\cite{jones2000software}.
Given the complexity and diversity of software development tasks, software tools are often designed to be highly versatile, with a huge selection of configuration options; some tools also support \emph{scripting}, which enables developers to construct complex customized usage scenarios.
Furthermore, some configuration options are based purely on user preferences (e.g., shell aliases, editor color scheme), thus no optimal configuration exists for all usage scenarios.
% properly introduce the term dotfiles.
As a convention, user-configuration files are often referred to as \dfs.
Detailed information regarding the history of the name \df can be find in \Cref{sec:background-dotfiles}.

% motivating example to show why concretely dotfiles are related to software engineering practices
To illustrate the importance of \dfs in software development, let us consider a motivating example of configuring ``hotkey'' commenting/uncommenting in the text editor \vim.
\Cref{listing:simple_vim_comment_toggles} presents a simple configuration that defines two ``hotkeys'' to allow \vim users to add or remove \emph{C}-style comments in source code with a single key-press.
%-- The configuration allows users to toggle between \Cref{listing:hello_c_before_commenting} and \Cref{listing:hello_c_after_commenting} by pressing the defined shortcuts rather than navigating through the user interface.
%--
\begin{figure}[ht]
\begin{lstlisting}
nnoremap <F2> :norm ^i// <C-[>
nnoremap <F3> :norm ^3x<C-[>
\end{lstlisting}
  \caption{Simple configuration for toggling comments in \vim}
  \label{listing:simple_vim_comment_toggles}
\end{figure}
%-- \noindent
%-- \begin{figure*}
%-- \begin{minipage}{.45\textwidth}
%--   \lstinputlisting[language=c,caption={Hello World in C},label={listing:hello_c_before_commenting}]{assets/hello.c}
%-- \end{minipage}\hfill
%-- \begin{minipage}{.45\textwidth}
%--   \lstinputlisting[language=c,caption={Hello World after pressing F2 on line 4},label={listing:hello_c_after_commenting}]{assets/hello_commented.c}
%-- \end{minipage}
%-- \end{figure*}

% sharing can lead to improved configuration, just like how open source development evolve in ESR's model
While this short configuration is functional, there is room for customization.
When the configuration is publicly available, through \github or a community wiki, other users may leverage the configuration as a base point and adapt it for improvement or other personalized use.
For example, another user may wish to change the configuration to use different hotkeys that they prefer; or, they may improve the script to support commenting in additional programming languages (e.g., \texttt{\#} in \emph{Bash}).
If these changes are shared back to the community, the author of the original configuration may adapt their own script to add the improvements.
As highlighted by a recent study that focuses on security issues in \dfs management~\cite{jungwirth2023connecting}, sharing is the top reason why developers host their \dotrepos publicly on \github followed by setup and backup.

% existing practice of sharing dotfiles
Although there is a vast body of empirical research on how software developers use their tools~\cite{johnson2013don,damevski2016mining,snipes2014experiences,murphy2019developers,kavaler2019tool}, at the same time there has been little study of how developers configure their tools and manage their configurations.
% \subsection{Goal of this paper}
In this work, we study \dfs that have been publicly shared on \github to better understand the practice of maintaining user-configuration files.
We collect the \dotrepos from \github using popularity and activity metrics with the help of \emph{GHTorrent}~\cite{gousios2012ghtorrent}.
Specifically, our study investigates three research questions:
% \newpage

{\noindent\bf \nameref{sec:rq1}}

Here, we study the prevalence of sharing \dfs among developers to better understand how widespread the practice is.
We wish to confirm if it is software developers --- rather than casual users --- who are doing the sharing, since \dfs can also be used for more general-purpose software used by the broader public.
To answer this RQ, we first traced the occupation of the owners of \dotrepos collected by their public profile; we observed that the majority of them appear to work in a field that involves programming activities.
We then checked the top 500 most-starred users on \github, and we found that 129 (i.e., 25.8\%) of these top users own a variant of a \dotrepo on \github.

{\noindent\bf \nameref{sec:rq2}}

In investigating this RQ, we hope to build an understanding of which specific configuration files developers manage and track with \dotrepos.
We extracted the \dfs by their normalized names (e.g., adjusted for different folder hierarchy structures) from the \dotrepos, and created a taxonomy of the top fifty most common \dfs in our dataset.
We find that configurations for text editors and \nix shells are most common in \dotrepos.
Meanwhile, meta-files such as \readme files, software license
information, and deployment scripts (e.g., \texttt{setup.sh}, \texttt{Makefile}) are also common.

{\noindent\bf \nameref{sec:rq3}}

In investigating this RQ, we aim to understand how developers maintain their \dfs.
We sampled 400 commits uniformly at random in \dotrepos, and performed an open card-sort to infer the intent of the commit.
We then selected the most frequently edited \dfs in all \dotrepos and modeled their code churn histories as time-series.
We used the state-of-the-art time-series clustering technique~\cite{paparrizos2015k} to extract patterns of code churn history of frequently edited \dfs.
We find that 54.8\% of commits directly change the configuration of software tools; meanwhile, other commits focus on managing the \dotrepo such as updating documentation and adjusting deployment scripts.
We observe that all types of \dfs can be found in every code churn history pattern, suggesting developers play a more important role than the type of \dfs in the frequency of editing \dfs.

The key contributions of this work includes (1) collecting a set of quality shared \dotrepos (available in replication package); (2) providing empirical evidence of the prevalence of \dfs sharing among developers; (3) providing a taxonomy of commonly shared \dfs; (4) identifying the intent of commits in \dotrepos; and (5) extracting code churn history patterns of frequently edited \dfs.

\section{Background}

In this section, we discuss the history of \dfs and the inception of sharing \dfs within the software development community.

\subsection{Dotfiles}\label{sec:background-dotfiles}

In this paper, we use to the term ``dotfiles'' to refer to a collection of user-specific configuration files~\cite{wiki.dotfiles}.
The term \dfs in its original meaning refers to the hidden files in UNIX-like (\nix) operating systems.
The concept originates from a bug in an early implementation of the command
\textit{ls}~\cite{dotfiles.history}, which produces a listing of the files in a given directory.
The bug occurred when the code --- correctly --- ignored the two special files that represent the current and parent directory (i.e., \texttt{"."} and \texttt{".."}), but also --- incorrectly --- ignored all other files that started with a dot.
Subsequently, a convention arose within the \nix community to prepend a period onto the beginning of file names that store application settings in users' home directories; these files would be ``hidden'' by default from the user's view unless they explicitly asked to see the \dfs, such as via the \texttt{-a} option to \texttt{ls}.
``Hiding'' the configuration files in this way reduces visible clutter, and reduces the risk of new users accidentally changing or removing settings files that they may not understand well.
With the widespread adoption of storing configurations in \dfs, the meaning of the term has since evolved to refer to the broader collection of user-specific configuration files.

As suggested by a community website dedicated to \dfs~\cite{dotfiles.github}, users can backup their \dfs online, learn from the existing \dfs from the broader community, and share back what they have learned to others. 
While the first occurrence of \dfs sharing is likely lost to history, the activity can be observed at least as far back as the early 1980s, when \emph{USENET} and e-mailing lists served as proto-social media for developers~\cite{emacs.mail.list}.
Users often share their personal configuration and wisdom for a program where others can benefit from; this culture of sharing has fostered a vibrant user community.
One blog post stood out that brought the idea to a larger community: Holman, one of the first engineers at \github, wrote a impactful blog entry titled ``Dotfiles are meant to be forked''~\cite{holman.blog}.
His own \dotrepo~\cite{holman.blog} is also one of the most starred \dotrepo on \github.
With \github gaining more popularity and becoming the largest source code hosting platform, many developers have chosen to host their \dotrepo on \github.

In our work, we leverage \github as the source to retrieve developers' \dfs repositories, and use these repositories as the foundation of our analysis.

\section{Data Collection}

Historically, managing \dotrepos is mostly an informal practice; as there is no universal method to manage a \dotrepo, it is impossible to extract all \dotrepos from \github.
We decided to perform an exact name match using the most common name ``dotfiles'' in the \emph{GHTorrent}~\cite{gousios2012ghtorrent} database from late 2019.
We ignored all forked repositories for the query.
We found \wzVTotalDotfilesRepos \dotrepos from the query.

Since \github is known to be filled with low-quality projects (e.g., temporary student projects that are unmaintained)~\cite{kalliamvakou2014promises}, we performed further filtering with three criteria to ensure the quality of the collected \dotrepos.
%-- We list the three requirements we made and also the reasoning behind the them below.
Specifically, we required that:

%-- \begin{enumerate}
%--   \item $\ge 5$ stars --- with endorsement from other users
%--   \item $\ge 5$ commits in the \dotrepo --- with active history
%--   \item $\ge 10$ commits in other repositories --- with active development history
%-- \end{enumerate}

\begin{enumerate}
\item each \dotrepo have been starred at least five times on \github (i.e., have explicit endorsements from other users),
\item each \dotrepo have had at least five commits made to it, and
\item the owner of each \dotrepo have made at least 10 commits to other \github repositories.
\end{enumerate}
Overall, we hope the collected \dotrepos will provide a basis for understanding how developers manage their \dfs; the filtering is intended to ensure that the \dotrepos we collect (a) have been found to be of use to others (b) have been actively maintained, and (c) belong to a software developer.
Since some \dfs do not necessarily correspond to development tools (e.g., configuration for window managers), we want to avoid \dotrepos that focus on these aspects (e.g., aesthetic customizations~\cite{unix.porn}).
In RQ1, we further explore on this aspect on the owner of the selected \dotrepos and also the popularity of owning \dotrepos among the most prolific \github users.

% -- the following is removed to save space.
% After the filtering process, we have \wzVFilteredDotfilesRepos \dotrepos.

We then continued to clone the \dotrepos from \github.
Due to the time difference between the \emph{GHTorrent} dataset (late 2019) and the time of cloning for our study (mid-2020), a small proportion of repositories were found to be no longer available; some common reasons for this includes migrating to another platform (e.g., \emph{GitLab}) and \emph{DMCA} take-down requests.
In the end, we are able to retrieve \wzVTotalClonedDotfilesRepos \dotrepos from \github out of the \wzVTotalDotfilesRepos from our original query.
Applying our three filtering criteria, the remainder resulted in a data set of \wzVDatasetDotfilesRepos.
We refer to the filtered \wzVDatasetDotfilesRepos \dotrepos as the \dotdataset.
To enable reproducibility and follow-up studies, we also share the data publicly\footnote{\url{https://zenodo.org/record/8368471}}.

\section{Results}

In this section, we answer the research questions proposed, and we make several observations about the sharing of \dotrepos.

\subsection{RQ1: Who are the owners of shared \dotrepos?}\label{sec:rq1}

\rheader{Motivation}
Sharing \dfs is often endorsed by the community~\cite{holman.blog} as a best-practice for software developers.
However, the scale of \df sharing in practice remains unclear.
Moreover, software development tools are not the only kind of applications that store configuration information in \dfs.
Many non-development-focused tools --- such as \nix \emph{PDF} readers --- also store their configuration in \dfs~\cite{unix.porn}.
While these applications are still important for developers, they do not focus directly on software development.
Thus, in this RQ, we aim to understand the popularity of sharing \dfs among developers, and evaluate our \dotdataset by verifying whether the \dotrepos are owned by developers.

We collected this information in Sept.\ 2022 leveraging a community tracker~\cite{github.most.stars}.
With the user ID collected, we used the \github API to retrieve the list of repositories owned by each user in our list.
For each user, we checked whether a \dotrepo exists by fuzzy matching the project names with ``dotfiles'' based on the Levenshtein distance~\cite{levenshtein1966binary}.
We went through all fuzzy matches with scores higher than 60 and selected the following two heuristics to only include repository names that are closely related to \dfs.
First, we use a score of 85 as the cutoff score where we would include repository names such as \texttt{dotfiles-local} and \texttt{myDotfiles}.
We then also require the project name to be at least four characters long, so repository names such as \texttt{fl} are ignored.
The two heuristics ensure the filtered repositories are mainly \dotrepos.

We then manually check the profession of the owners of the popular \dotrepos we collected to verify if our filtering process successfully yields \dotrepos that belong to developers.
We sampled 100 \dotrepos uniformly at random from the \dotdataset to try to infer the profession of the owner.
For each repository, we start by visiting the owner's \github profile homepage.
When the user provides additional information such as a personal profile, personal website links, and \emph{LinkedIn} profile page, we followed the publicly available information and took note on the owner's stated profession, if we could find it.
When the information was not present, we supplemented the process with a simple web query based on the their \github user name.
In these cases, we can sometimes discover the owner's information as people tend to use the same user name across platforms.
We restrained ourselves from further investigations (e.g., leveraging the author name and e-mail) to avoid violating the users' privacy.

\begin{figure}
  \centering
  \includegraphics[width=0.45\textwidth]{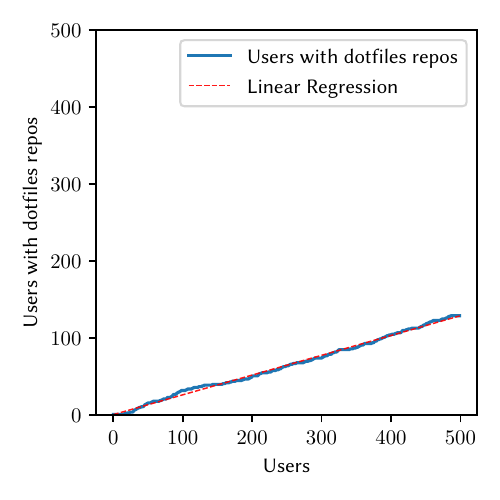}
  \caption{Number of top users by total repo stars on \github with \dotrepos. The dotted line suggests that the percentage is stable as we consider more users.}
\label{fig:users_with_dotrepos}
\end{figure}

\rheader{Results}
We now report our findings from investigating the most-starred users on \github, and from tracking the public profiles of the owners of \dotrepos contained in the \dotdataset.
%-- We start by making the following observation:

\observe{Developers often share \dfs.}
Owning a \dotrepo is common among the most prolific developers on \github.
We found that \wzVTopUsersWithDotfiles (i.e., 25.8\%) of the top 500 most-starred users have a \dotrepo where \wzVTopUsersWithDotfilesStrict of them are named exactly ``dotfiles''.
We note that the actual number may be higher than the measured one, since the ``user'' here may represent a organization instead of a single developer.\footnote{\url{https://docs.github.com/en/get-started/learning-about-github/types-of-github-accounts}}
In \Cref{fig:users_with_dotrepos}, the solid blue line represents the cumulative number of users with a \dotrepo in the top 500 most-starred users.
That is, at the 100 mark on the x-axis, it represents the number of users with a \dotrepo in the top 100 most-starred users.
As shown in the figure by the dotted red line, we observe a linear growth in the number, suggesting the prevalence of sharing \dfs among the most-starred users of \github.

We found that the majority of the owners of \dotrepos in the \dotdataset appear to work in a programming-related job.
As shown in \Cref{tab:dotfiles_owner_jobs}, 79 \dotrepos owners are software developers, with 3 system admins, 6 students, and 3 researchers.
Unfortunately, we were unable to identify the profession for 9 of the owners.
This finding gives us high confidence that the \dfs contained in the \dotdataset are mostly from developers.

\begin{table}[htbp]
\centering
\caption{Sampled 100 \dotrepos owners}
\label{tab:dotfiles_owner_jobs}
\begin{tabular}{@{}lr@{}}
\toprule
Self-declared & \# of owners of \\
profession & \dotrepos \\
\midrule
Software developer                & 79                        \\
System admin             & 3                         \\
Student                  & 6                         \\
Researcher               & 3                         \\
Unknown                  & 9                         \\
\bottomrule
\end{tabular}
\end{table}

We believe the prevalence of sharing \dfs comes from the familiarity that developers have during their regular job to use version control systems.
Like source code, the inherent plain text nature of \dfs makes their tracking equivalent to tracking source code.
Consequently, developers can effortlessly manage the evolution of their \dfs using the same methods employed for managing source code.
%-- Similar to source code, the plain text property of \dfs makes tracking them the same as tracking source code.
%-- As a result, developers can easily manage their \dfs in the same way they manage source code.

\subsection{RQ2: What kind of user-specific configuration files do users track in their \dotrepos?}\label{sec:rq2}

\rheader{Motivation}
After confirming that our \dotdataset originates mainly from software developers, in this RQ we explore the content contained in \dotrepos.
Since \dfs are user-specific configurations, we first aim to identify which software tools developers are explicitly maintaining configuration settings for.
With firm evidence of what specific information developers track in \dotrepos, we can better understand the tools that are likely to be customized.

\rheader{Methodology}
We extracted all unique \dfs from the \dotdataset.
We represent each \df by its lowercase base name, and remove the leading dot if any.
For example, a file \texttt{vim/.VIMRC} (relative to \emph{git} base) will be normalized to \texttt{vimrc}.
We applied this step since the structure of each \dotrepo is unique.
Developers may have different approaches to managing \dfs (e.g., writing deployment scripts, using \dfs management tools), so the normalization step ensures that we are correctly tracking the \dfs across different structures.
We also ignore the file extension for \readme files to accommodate the developers' choice of format; thus, \readme files in \emph{Markdown}, \emph{Org}, \emph{Asciidoc}, etc.\ are treated uniformly.
After normalizing the \df names, we create a taxonomy of \dfs based on the top 50 most common \dfs by percentage of presence.
The taxonomy is based of the type and purpose of the application each \df belongs to.

\begin{figure}
  \centering
  \begin{subfigure}[b]{0.45\textwidth}
    \centering
    \includegraphics[width=\textwidth]{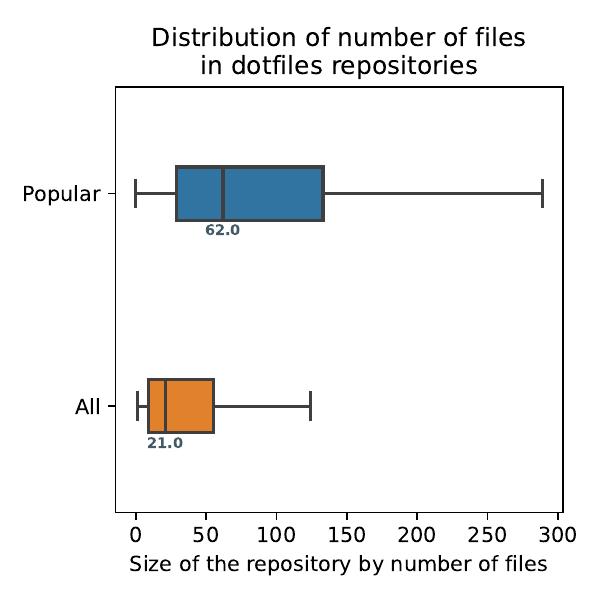}
    \caption{Number of files in \dotrepos}
    \label{fig:dist_n_files}
  \end{subfigure}
  \hfill
  \begin{subfigure}[b]{0.45\textwidth}
    \centering
    \includegraphics[width=\textwidth]{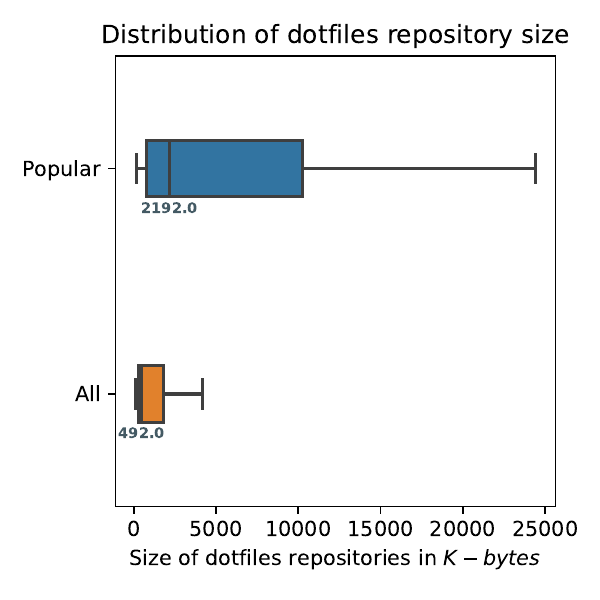}
    \caption{Size of \dotrepos}
    \label{fig:dist_raw_size}
  \end{subfigure}
  \caption{Content size of \dotrepos}
  \label{fig:stats_on_dotfiles_repos}
\end{figure}

\rheader{Results} \Cref{fig:stats_on_dotfiles_repos} provides a overview of the size of \dotrepos.
There are a median of 62 files in \dotrepos and they take up a median about 2Mb of space.

In \Cref{tab:pop_dotfiles}, we list the top 20 with the percentage of \dotrepos they are present in the \dotdataset.
In addition to project meta files (e.g., \readme, \emph{license}), we found that configuration files for \vim, \git, \emph{tmux}, \emph{zsh}, and \emph{bash} to be the most common.

\begin{table*}[htbp]
\centering
\caption{20 most tracked \emph{dotfiles} by popularity}
\label{tab:pop_dotfiles}

\begin{tabular}{@{}lllr@{}}
\toprule
%-- Filename   & Associated & Type & \#(\%) \\
%-- 	    & software & \\
Filename   & Associated application & Application type & \#(\%) \\
\midrule
    \texttt{README}           &              & meta                 & 2923 (88.4\%) \\
    \texttt{gitignore}        & \emph{git}      & version control      & 2672 (80.8\%) \\
    \texttt{vimrc}            & \emph{vim}      & text editor          & 2113 (63.9\%) \\
    \texttt{gitconfig}        & \emph{git}      & version control      & 2087 (63.1\%) \\
    \texttt{tmux.conf}        & \emph{tmux}     & terminal multiplexer & 1950 (59.0\%) \\
    \texttt{zshrc}            & \emph{zsh}      & shell                & 1856 (56.2\%) \\
    \texttt{config}           & multi\emph{*}   &                      & 1486 (45.0\%) \\
    \texttt{bashrc}           & \emph{bash}     & shell                & 1342 (40.6\%) \\
    \texttt{gitmodules}       & \emph{git}      & version control      & 1131 (34.2\%) \\
    \texttt{bash\_profile}     & \emph{bash}     & shell                & 954 (28.9\%)  \\
    \texttt{init.vim}         & \emph{vim}      & text editor          & 904 (27.4\%)  \\
    \texttt{license}          &             & meta                 & 844 (25.5\%)  \\
    \texttt{inputrc}          & \emph{readline} & text edit            & 798 (24.1\%)  \\
    \texttt{xresources}       & \emph{Xorg}     & display server       & 709 (21.5\%)  \\
    \texttt{install.sh}       &             & meta                 & 698 (21.1\%)  \\
    \texttt{gemrc}            & \emph{Ruby}     & package manager & 625 (18.9\%)  \\
    \texttt{xinitrc}          & \emph{xinit}    & start \emph{Xorg}    & 601 (18.2\%)  \\
    \texttt{gitignore\_global} & \emph{git}      & version control      & 537 (16.2\%)  \\
    \texttt{zshenv}           & \emph{zsh}      & shell                & 523 (15.8\%)  \\
    \texttt{aliases}          & shell           & shell                & 510 (15.4\%)  \\
\bottomrule
\end{tabular}
\\
\bigskip
\raggedright
\emph{*:} \texttt{config} is a common name used by multiple software developers as their default configuration file name.
Some common examples include \emph{i3wm} a window manager, \texttt{$\sim$/.config/i3/config}, and \emph{fcitx5} a input method manager, \texttt{$\sim$/.config/fcitx5/config}.

\end{table*}

\observe{Configurations for shell, text editors and \git meta files are the most common \dfs.}

We observe that configurations for \nix shells --- specifically, \emph{zsh}, \emph{bash}, and \emph{fish} --- also contribute to the top 50 list.
%--We observe configuration files associated with three different shells, namely \emph{zsh}, \emph{bash}, and \emph{fish}.
Auxiliary files for shells such as \texttt{aliases}, \texttt{profile}, and \texttt{zshenv} are also common; these files complement the shell configurations by splitting the configuration based on functionality (e.g., separate aliases) and environment (e.g., login and interactive sessions).

\begin{figure*}
  \includegraphics[width=\textwidth]{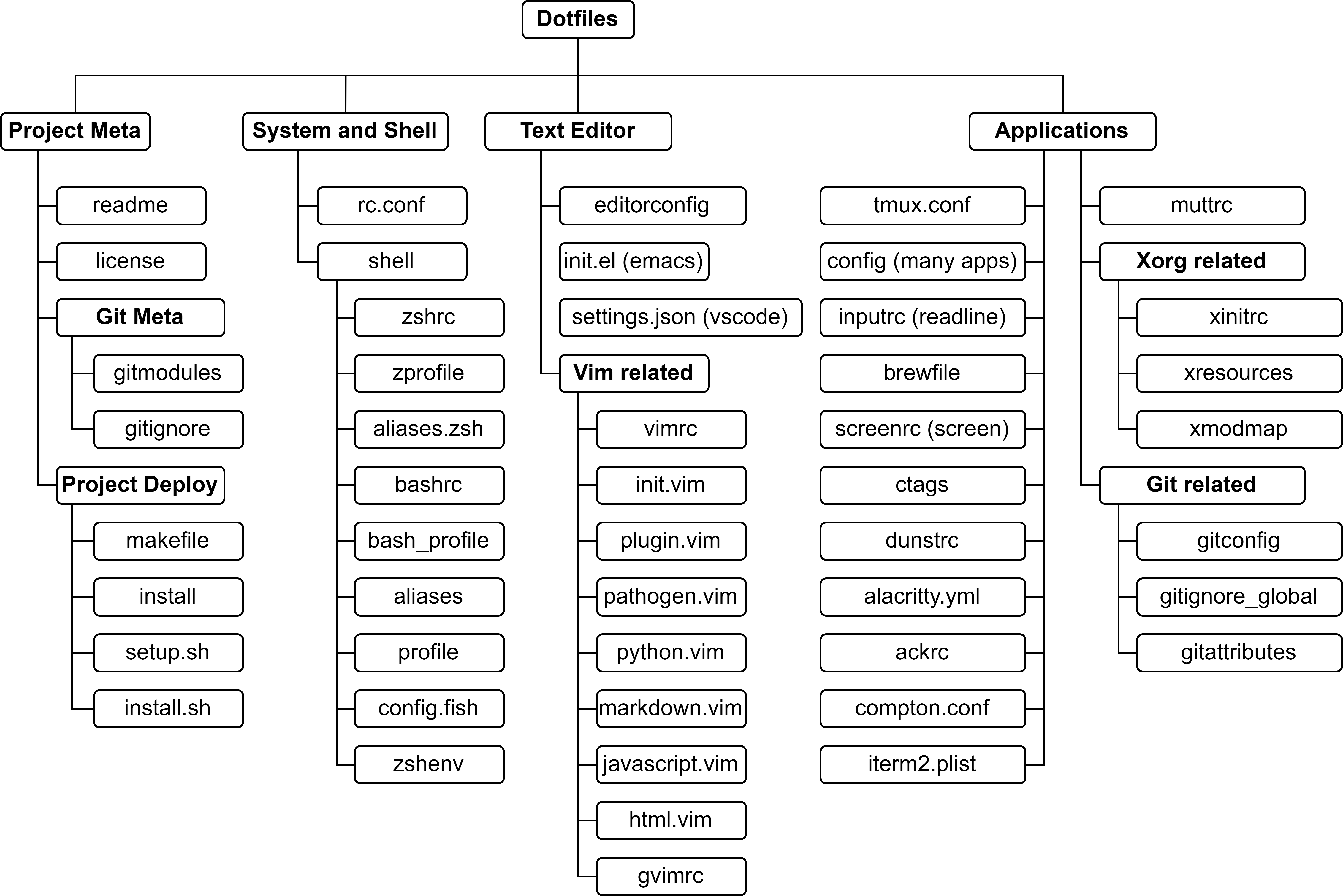}
    \caption{Taxonomy of the top 50 most common \dfs}
\label{fig:dotfiles_taxonomy}
\end{figure*}

Another common file type in the top 50 list are text editor-related configurations, specifically those for \vim, \emacs, \vscode, and \emph{editorconfig} which is respected by multiple editors.
\vim allows file-type-specific setups in separate files, which explains the 6 different \texttt{.vim} files associated with different languages.
The remaining 3 files correspond to variants of \vim, the default (\texttt{.vimrc}), the graphic version of \vim~--- \emph{GVim} (\texttt{gvimrc}), and a major fork of \vim~--- \emph{NeoVim} (\texttt{init.vim}).
We note that despite \vscode being the most popular text editor \cite{so.survey.2022}, the popularity of \texttt{settings.json} is overshadowed by \vim in our study.
We conjecture that \vscode encourages users to use its builtin synchronization support, removing the needs for managing the configuration through a \dotrepo.

Finally, we observed that \git meta files were present in our top 50 list; we found this to be unsurprising since the \dotrepos are managed by \git.

\observe{Meta-files are common in \dotrepos.}
We also note the presence of \dotrepo-specific files in the top 50 list.
Commonly found in \github projects, \readme and \texttt{license} are also common in \dotrepos.
The \readme file often contains information about the repository and instructions to deploy the \dfs in a new environment.
The detailed deployment method can leverage an existing \dfs management system, such as \emph{Chezmoi}~\footnote{\url{https://github.com/twpayne/chezmoi}} or \emph{GNU stow}, or with custom install scripts as \texttt{install.sh} or \texttt{Makefile}, both of which are present in the top 50 list.

\observe{GUI configurations are rare in \dotrepos.}
In \Cref{fig:dotfiles_taxonomy}, we show a taxonomy of the 50 most popular \dfs.
We can see immediately that most \dfs are for command-line applications, while only a few \dfs are for \emph{GUI} application.
For example, \texttt{iterm2.plist} for \emph{iTerm} on \emph{MacOS}, and \texttt{dunstrc} for the notification manager \emph{dunst};
there are also a few files that relate to the \emph{Xorg} display server, such as \texttt{xinitrc} and \texttt{xinit}, and the \emph{GUI}-specific \vim configuration \texttt{gvimrc} is also present in the list.
We suspect the lack of \emph{GUI} configuration files is because --- unlike in the 1980s \nix world --- modern \emph{GUI} applications are often configured using a \emph{GUI} interface, and those settings are managed and stored implicitly by the application rather than explicitly by the user.
While the \emph{GUI} apps offer a simpler way to tweak the configurations, it also makes sharing and replicating the configurations a more manually focused process.
Our observations are based on the \dfs we analyzed and the comparison of \emph{GUI} and \emph{TUI} applications is out of the scope of this paper.

Some applications also have their own mechanism for managing configuration (e.g., web browsers).
We note that despite \emph{Windows} being common for development~\cite{so.survey.2022}, we did not observe any \emph{Windows}-specific software configuration files in the top 50 list.
We observed some instances of \emph{AutoHotKey} scripts which is a popular Windows application for automation.
We suspect the lack of \emph{Windows}-specific configuration files is due to \emph{Windows} prefer using the \emph{registry} database in which applications and system components store and retrieve configuration data~\cite{windows.registry}.

\subsection{RQ3: How do developers update their \dfs?}\label{sec:rq3}

\rheader{Motivation}
We theorize that developers often update their \dfs to adjust the configurations to adapt to both the change in development need (e.g., switching to a new language) and also the change in personal taste (e.g., testing out alternative tools).
The \dotdataset, which contains real-world \dfs, allows us to investigate the details of how developers update their \dfs.

Through studying the change history of \dfs, we can build an understanding of how developers maintain the \dotrepo.
Also, through studying the frequently edited \dfs, we can better grasp the patterns of how developers update the \dfs.
Our overall theory is that different types of \dfs may be updated differently.
For example, after a rapid period of changing the \readme file, the information will stabilize and require only infrequent updates subsequently.
Meanwhile, for other tools such as \emacs, the developers may fine tune it frequently to adapt the tool to their needs.

In this RQ, we study the intent of \dotrepo commits and the code churn history patterns to better understand how developers update their \dfs.

\rheader{Methodology}
We start by sampling 400 commits from the \dotdataset.
The two authors then performed open card-sorting on the commits to derive the intent of the commit.
For each of the 400 commits, we printed what we can fit on a piece of paper.
Most of the commits take less than half of the piece of paper, and in case of very long commits, we referred to the full content on our computers.
We went through the pile of paper and examined each commit to group similar commits together.
We iterated through the groups multiple times until the groups are finalized and all authors agree with the results.
This step allows us to build an understanding of why \dfs are updated by developers.

With the next step, we attempt to discover historical patterns in code churn.
We tracked the total number of commits for each \df in the \dotrepo.
We extracted frequently-edited \dfs that have at least 20 commits and are at least 1 year old.
We modeled each of the frequently updated \dfs as a time-series using its code churn history.
The time-series is represented by cumulative total code churn summed in each commit (i.e., sum of added and deleted lines).
Since most time-series analysis techniques require regular time-series with the same number of timesteps,
we limit the time-series from the last step to only the first year, and normalize the timesteps to 1 day.
At this step, each time-series have 365 data points.
We then removed time-series which have fewer than 20 updates.
In other words, the resulting time-series are from \dfs that have at least one year of history and
have at least ten commits associated with it from ten different days in the first year.
In the end, we have a total of \wzVTSCount time-series representing \wzVTSUniqueCount distinct \dfs in the frequently updated set of \dfs.

% Processing of applying k-shape
We then leverage a time-series clustering method called \kshape~\cite{paparrizos2015k} to extract the patterns of code churn history.
We first normalize the time-series with a mean-variance filter; this filter allows us to remove the bias from absolute change of the time-series and focus more on the relative change.
For example, when two files share the percentage of change (e.g., 10\%) while one file is 500 lines and the other is 50, the mean-variance filter is able to normalize the two time-series.

We then leveraged the \kshape clustering algorithm for time-series clustering.
The \kshape algorithm uses metrics that focus on the shape of the time-series.
This approach has been found to have similar performance to \emph{DTW}~\cite{sakoe1971dynamic}, which, in turn, has been shown to outperform Euclidean distance, and is more computationally efficient than \emph{DTW}~\cite{gorecki2019comprehensive}.

% How we decided on the k-value
\kshape, which is derived from \emph{K-means}~\cite{macqueen1967some}, also suffers from the problem of having to determine an appropriate value for $k$.
Since the distance between time-series, and determining the centroid of a series of time-series is still an on-going challenge in time-series analysis,
there is no universal method to validate the result of the clusters~\cite{aghabozorgi2015time}.
So we experimented with different $k$ values ranging from $2$ to $10$,
and selected the best $k$ value (i.e., 4) based on our interpretation of the clustering results.

\begin{figure*}
  \centering
  \begin{subfigure}[b]{0.30\textwidth}
    \centering
    \includegraphics[width=\textwidth]{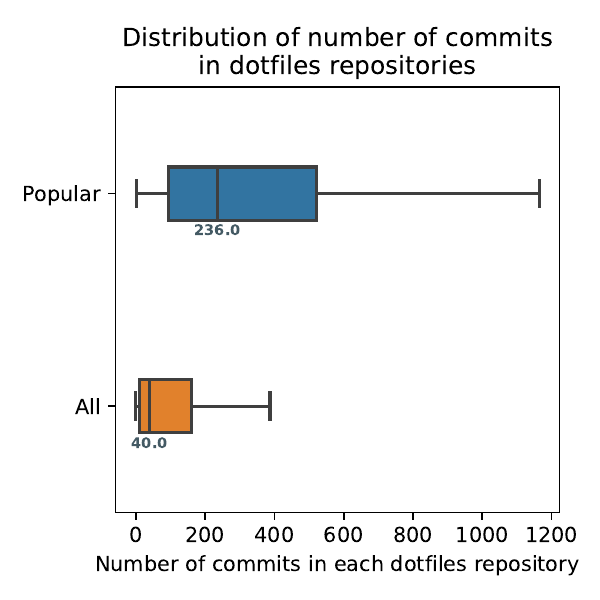}
    \caption{Distribution of \dotrepos commits}
    \label{fig:dist_n_commits}
  \end{subfigure}
  \hfill
  \begin{subfigure}[b]{0.30\textwidth}
    \centering
    \includegraphics[width=\textwidth]{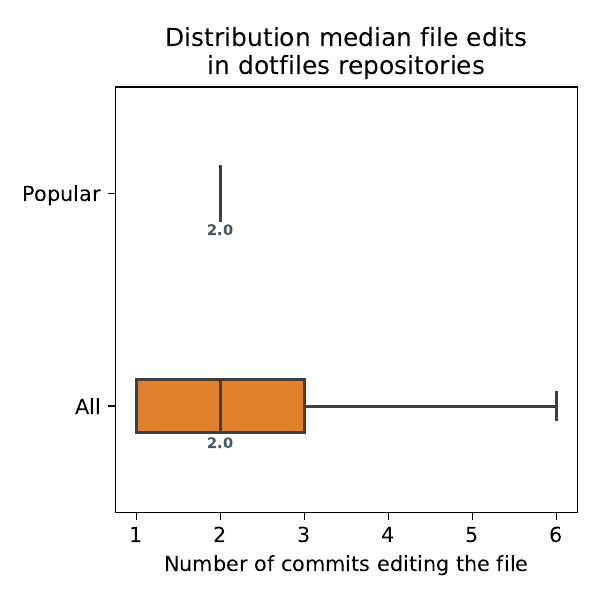}
    \caption{Distribution of \dotrepos median file updates}
    \label{fig:dist_median_file_edits}
  \end{subfigure}
  \hfill
  \begin{subfigure}[b]{0.30\textwidth}
    \centering
    \includegraphics[width=\textwidth]{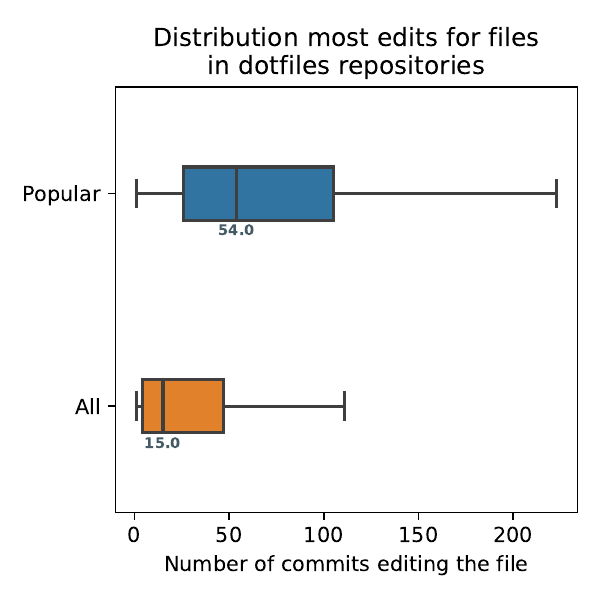}
    \caption{Distribution of \dotrepos most updated file}
    \label{fig:dist_most_file_edits}
  \end{subfigure}
  \caption{Information on commits for \dotrepos}
  \label{fig:stats_on_commits}
\end{figure*}

\rheader{Results}
In \Cref{fig:stats_on_commits}, we show the distribution of the number of commits, the median commits by file per \dotrepo, and the number of commits for the most edited file per \dotrepo.
As suggested by \Cref{fig:dist_n_commits}, the \dotrepos are updated frequently (note that in our filtering process, we require the \dotrepo to have only five commits).
We can observe a similar trend, where the popular \dotrepos from \dotdataset have more commits and contain files that are edited more often as suggested by the max file edits.
However, the median file edits is same for both \dotdataset and all \dotrepos, suggesting that despite the activity difference, most files in \dotrepos remains unchanged.

\observe{The majority of \dfs receive at most one update.}
%-- \obs{The majority of \dfs never or only receive one update.}
As shown in \Cref{fig:dist_median_file_edits}, in most \dotrepos, the median number of commits per file is only two.
Since the \df is introduced to the repository in the first commit, it means that the \df have only received one update since its introduction.
This suggests that a large proportion of the \dotrepo focuses on ``cold storage'' where the content rarely gets updated.

\observe{Most \dfs updates focus on a small set of files.}
%-- \obs{Most \dfs updates focus on a small set of files.}
Although most \dfs seldom receive updates, the most frequently updated \df in each \dotrepos receive many updates as shown in \Cref{fig:dist_most_file_edits}.
While most \dfs remain stable, developers update some \dfs frequently.

\observe{Most \dotrepo commits relate to tweaking the behavior of software tools.}
We found that the majority (i.e., 63.3\%) of the \dotrepo commits relate to minor behavioral ``tweaking'' of the software tools.
We also found that 30.8\% of the commits edited configurations in the form of scripting (e.g., improving a shell script),
24.0\% of the commits changed parameterization options (i.e., a predefined configuration option), and 8.5\% of the commits relate to creating and modifying shortcuts (e.g., adding a \git alias).
The details of the card-sorting results are shown in \Cref{tab:dot_commits}.
Scripting is often related to defining custom actions in configuring software; it can be most commonly found in shell scripts for automation (e.g., swapping the location of two files) and performing complex functionalities in text editors (e.g., run code formatters when a file is saved).

\begin{table}[htbp]
\centering
\caption{Type of \dfs commits}
\label{tab:dot_commits}

\begin{tabular}{@{}lr@{}}
\toprule
Type                           & \#(\%)       \\
\midrule
tweaking script behavior       & 123 (30.8\%) \\
tweaking parametrized options  & 96 (24.0\%)  \\
dotfiles deployment management & 41 (10.3\%)  \\
fixing bugs in configuration   & 37 (9.3\%)   \\
tweaking shortcuts             & 34 (8.5\%)   \\
refactoring configuration      & 17 (4.3\%)   \\
managing external resources    & 23 (5.8\%)   \\
documentation update           & 20 (5.0\%)   \\
misc                           & 18 (4.5\%)   \\
\bottomrule
\end{tabular}
\end{table}

\observe{A significant amount of commits focus on \dfs management.}
We found that a significant amount of commits (i.e., 25.4\%) focus on \dfs management.
Documentation updates occur in 5.0\% of the commits where developers update the contents of \readme files.
10.3\% of the commits directly involve changing the files related to \dfs deployment.
These commits modify deployment specific files such as \texttt{Makefile} and \texttt{setup.sh}.
In 4.3\% of the commits, developers refactor their \dfs.
This is often indicated by the commit message such as ``clean up''.
We also observe 5.8\% of the commits that deal with external resources.
For example, some developers manage their \vim plugins with \git submodules leveraging package managers like pathogen~\footnote{\url{https://github.com/tpope/vim-pathogen}}).

The rest of the commits (i.e., 4.5\%) do not fit in either of the above mentioned categories.
These commits are often aggregations of multiple changes and do not have a single purpose.

% k-shape

\begin{figure*}
    \includegraphics[width=\textwidth]{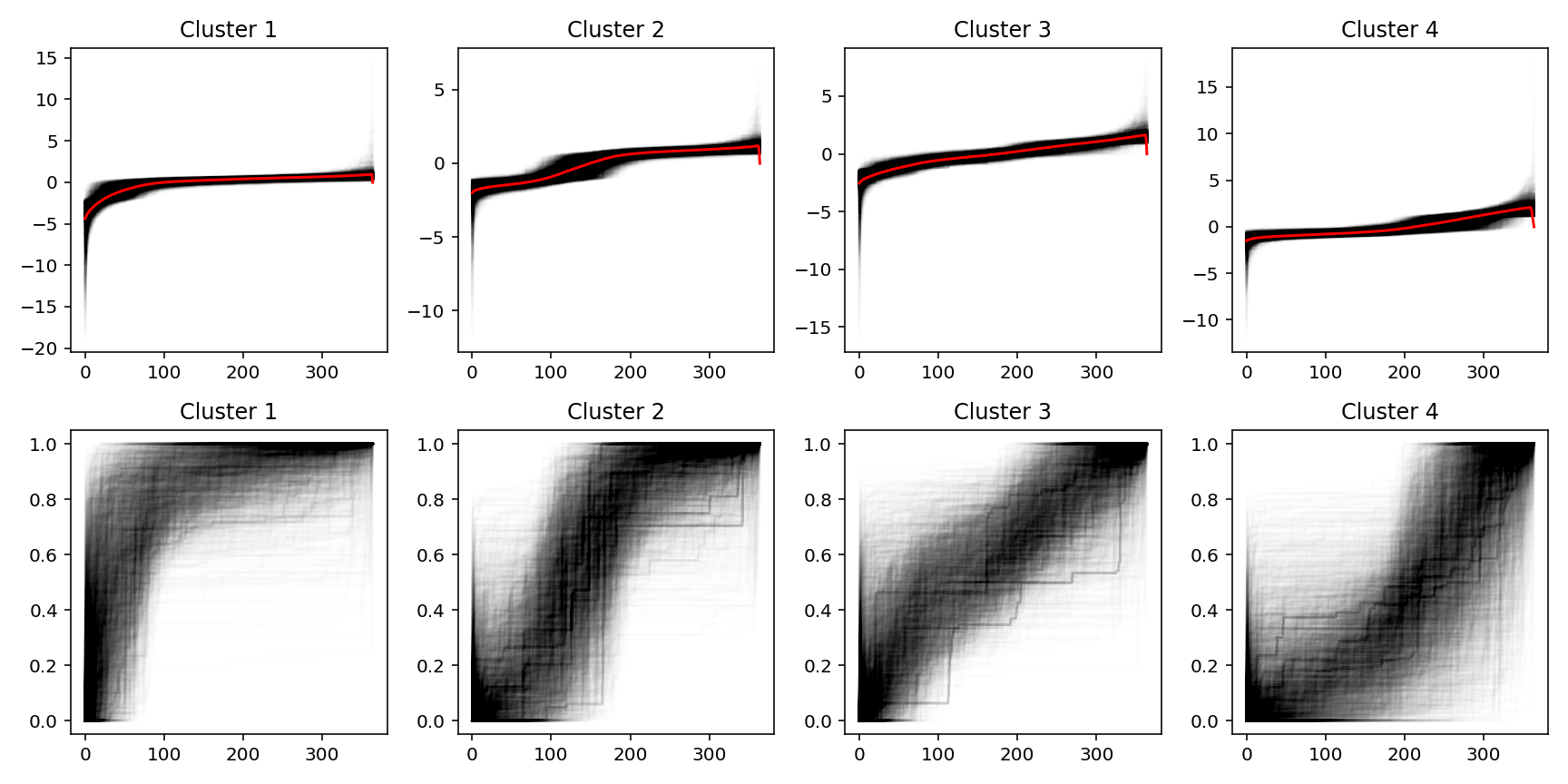}
    \caption{\kshape clustering ($k = 4$) results on time-series modeled by code-churn history. Top: Mean-variance normalized. Bottom: Min-max normalized.}
\label{fig:kshape_result}
\end{figure*}

% Describing the clusters

\observe{Three types of \dfs code churn history patterns can be observed in the $4$ clusters.}
In \Cref{fig:kshape_result},
we show the clustering results from \kshape in both mean-variance and min-max normalized form.
The figures on the top are the mean-variance normalized time-series.
These time-series are the raw data fed to \kshape.
The red line is the average of the time-series in each cluster.
The slope of the line after mean-variance normalization represents the change.
So when the slope is near zero it means that there is limited code churn during the period.
The figures on the bottom are min-max normalized between 0 and 1 for better visualize the code churn history.
Cluster~1 are \dfs that receive updates frequently after the \df is introduced to the \dotrepo, but remain mostly unchanged since then.
Clusters~2 and 4 are similar in the sense that these \dfs receive updates over time, but differ when the majority of the updates occur.
\dfs in Cluster~3 are continually updated over time and the modified size are consistent throughout.

% High level cluster differences
We also report here over observations on the clustering results in different $k$ values.
We settled on $k=4$ since it showed the best separation and consistency of the time-series.
When we began with $k=2$, it was immediately clear that one type of code churn history pattern is the \dfs stops receive updates shortly after it has been introduced.
As we gradually increase the value of $k$, we observe different patterns of when the updates occur over time.
Similar to Cluster~2, additional clusters in higher $k$ value will have a sudden update in different time range.
In other words, these clusters will overlap if we shift the time-series.

\begin{table}[htbp]
\centering
\small
\caption{Distribution of frequently edited \dfs across clusters (C)}
\label{tab:cluster_dist}
\begin{tabular}{@{}lrrrrrrrrr@{}}
\toprule
\df type      & C 1 \# &     \%    & C 2 \# &      \%   & C 3 \# & \%        & C 4 \# &  \%       & Total \\
\midrule
       \texttt{vimrc}  & 243       & 20.59\% & 325       & 27.54\% & 245       & 20.76\% & 367       & 31.10\% & 1180  \\
       \texttt{zshrc}  & 165       & 21.77\% & 194       & 25.59\% & 184       & 24.27\% & 215       & 28.36\% & 758   \\
      \texttt{readme}  & 63        & 17.50\% & 145       & 40.28\% & 69        & 19.17\% & 83        & 23.06\% & 360   \\
      \texttt{config}  & 95        & 26.61\% & 102       & 28.57\% & 75        & 21.01\% & 85        & 23.81\% & 357   \\
      \texttt{bashrc}  & 60        & 17.80\% & 103       & 30.56\% & 73        & 21.66\% & 101       & 29.97\% & 337   \\
    \texttt{gitconfig}  & 55        & 17.80\% & 73        & 23.62\% & 79        & 25.57\% & 102       & 33.01\% & 309   \\
    \texttt{tmux.conf}  & 59        & 19.60\% & 94        & 31.23\% & 65        & 21.59\% & 83        & 27.57\% & 301   \\
    \texttt{gitmodules}  & 42        & 14.63\% & 93        & 32.40\% & 70        & 24.39\% & 82        & 28.57\% & 287   \\
     \texttt{aliases}  & 37        & 17.21\% & 58        & 26.98\% & 41        & 19.07\% & 79        & 36.74\% & 215   \\
    \texttt{init.vim}  & 47        & 25.54\% & 49        & 26.63\% & 46        & 25.00\% & 42        & 22.83\% & 184   \\
    \texttt{install.sh}  & 37        & 26.81\% & 37        & 26.81\% & 27        & 19.57\% & 37        & 26.81\% & 138   \\
     \texttt{init.el}  & 34        & 25.00\% & 34        & 25.00\% & 20        & 14.71\% & 48        & 35.29\% & 136   \\
    \texttt{brewfile}  & 29        & 22.14\% & 36        & 27.48\% & 32        & 24.43\% & 34        & 25.95\% & 131   \\
    \texttt{bash\_profile} & 22        & 17.19\% & 18        & 14.06\% & 23        & 17.97\% & 65        & 50.78\% & 128   \\
    \texttt{gitignore}  & 32        & 25.20\% & 41        & 32.28\% & 15        & 11.81\% & 39        & 30.71\% & 127   \\
    \texttt{package.json}  & 8         & 7.14\%  & 28        & 25.00\% & 43        & 38.39\% & 33        & 29.46\% & 112   \\
     \texttt{xinitrc}  & 15        & 15.96\% & 22        & 23.40\% & 29        & 30.85\% & 28        & 29.79\% & 94    \\
    \texttt{aliases.zsh}  & 25        & 27.47\% & 21        & 23.08\% & 24        & 26.37\% & 21        & 23.08\% & 91    \\
    \texttt{plugins.vim}  & 16        & 20.25\% & 19        & 24.05\% & 22        & 27.85\% & 22        & 27.85\% & 79    \\
    \texttt{makefile}  & 12        & 15.38\% & 23        & 29.49\% & 20        & 25.64\% & 23        & 29.49\% & 78    \\
\bottomrule
\end{tabular}

\end{table}

% Overall distribution
The distribution of the top 20 most frequently updated files across the clusters is shown in \Cref{tab:cluster_dist}.
It is immediately clear that all types of \dfs are well represented in each of the clusters.
This suggests that the type of \df does not have a large effect on how the \df is maintained.
We believe the developers have a larger impact on how the \dfs are updated.
We checked the frequently updated \dfs by each developer and observe a median 57.1\% of the developer's frequently updated \dfs belong to one pattern.
In other words, if the developer has no impact, this percentage will be around 25\%.
However, note that the median number of frequently edited \dfs by developer is only three, so the data is likely biased.
A developer who enjoys tinkering may constantly update their \dfs to change their workflow.
At the same time, a goal-oriented developer may spend some time to configure the tools at first to get it working and stick to the configuration afterwards.

\section{Discussions and Implications}

Through our study of \dotrepos, we have gained insights in how developers configure their software tools and how they manage the configurations in \dotrepos.
Based on our observations, we discuss the challenges we perceive in managing \dotrepos.
Moreover, we discuss how shared collections of real-world user-specific configurations can potentially benefit developers, tool designers, and researchers.

\subsection{Challenges in Managing \dfs}

\challenge{Deployment of \dfs requires effort.}
For the \dfs to be read correctly by their associated software applications, the \dfs need to be stored at the correct location.
Common locations include the user's home directory and the \texttt{\$XDG\_CONFIG\_HOME} (which defaults to \texttt{\$HOME/.config}).
A typical application will have its own folder inside \texttt{\$XDG\_CONFIG\_HOME} often named after itself with the configuration files.
Sometimes the application's configuration file will also be called \texttt{config}.
Users often have some kind automated setup for moving the \dfs to the targeted location.
The setup can be through a deployment script (e.g., \texttt{setup.sh}), or through dedicated tools such as \emph{GNU Stow}.~\cite{gnu.stow}

In RQ2, we find many files among the most popular tracked \dfs that focus on automating the process of deploying \dfs.
The deployment process can be either fully automated, where everything is taken care of by a script, or in documentation that contains the steps for deployment.
Moreover, in RQ3, 10.3\% of the commits deal with managing the deployment of \dfs.
At the current stage, no universal method or tool exists for developers to manage their \dfs.
Similar observations can also be found in a recent survey~\cite{jungwirth2023connecting}, where they observed that most developers either used plain \git or no tools to manage their \dotrepo.

The lack of a standard method for managing \dfs can introduces challenges in sharing \dfs.
For example, when a developer wishes to explore other developers' \dfs, the developer must consult additional scripting and documentation to understand the setup of other developers.
Future research can investigate this aspect further through developer interviews and surveys to better understand the challenges in managing \dfs and develop methodologies and tools to improve the process.

\challenge{\dfs need to manage external resources.}
Similar to software development, we find that developers also rely on external resources in \dfs; for example, \emph{git submodules} can be used to include external dependencies in \dfs.
The external resources are plugins that extends the functionality of the software tools.
The most common type of external resources we encountered in our study are \vim plugins; we believe the popularity of this approach comes largely from the \vim plugin managers \emph{Vundle} and \emph{pathogen}.

External resources can also be involved indirectly.
Using \vim as another example, other plugin managers (e.g., \emph{vim-plug}), instead of using \emph{git submodules} requires only a declaration of the external resource by its URL.
However, unlike \emph{git submodules}, there could be challenges in \dfs replication this way.
For example, the version maybe different based on the time of retrieving the external resources.
We also observed developers trying to mitigate this issue by adding lock files that tracks the exact commits of the external resources.

\challenge{\dfs can leak privacy information.}
Some configurations may need to deal with sensitive information.
Jungwirth \etal{}~\cite{jungwirth2023connecting} studied the exposure of secrets in \dotrepos in depth, where they find potentially sensitive information is leaked by 73.6\% of \dotrepos.
They find that e-mail addresses are the most common type of exposed sensitive information, along with RSA keys and API keys.
%- Other types of sensitive information includes RSA keys and API keys.
%- in 73.6\% of \dotrepos, potentially sensitive information is leaked with e-mail addresses as the most common type of information.
%- Other types of sensitive information includes RSA keys and API keys.
As pointed out by a recent study, security leakage can be a huge concern on \github due to accidentally committing confidential information~\cite{feng2022automated}.
In our sample of commits studies, we observed developers taking actions to mitigate this issue.
For example, through using local environment variable, the developer can avoid storing sensitive information in plain text in the configuration files.

%
% =============================================================
%
\subsection{Leveraging \dfs as a Software Repository}

% \paragraph{The set of \dfs forms a software repository of practical user scenarios.}
One of the challenges of tool developers is to understand the user requirements.
However, given the complexity and customizability of software tools, the requirements may be complex and not single purpose.
One well known example is the concept of a bug becoming a feature.
A tool may be used differently from how it was originally designed, as unforeseen emergent uses become apparent to the user community.
We believe by leveraging the collection of \dfs, we can begin to address the challenges.

\opportunity{Analyzing \dfs to provide indirect user usage data.}
The rich real-world use case information contained in \dotrepos provides valuable information to understanding software usage.
We have observed many successes in leveraging telemetry-like techniques to better understand user behavior and in return improve user experience in software products~\cite{zheng2011massconf,raychev2014code}.
However, telemetry has been very challenging in the open-source community.
It is very hard to embed transparent telemetry in open-source software due to privacy concerns.
Most open-source software that implement some form of telemetry adopt an \emph{opt-in} philosophy, which most users do not actively turn on.
Traditionally, we rely on active members who participate and contribute to discussions to move forward on user experience.
And from time to time, we can observe patters of ``scratching an itch'' type of contribution made by other users.
However, this means that problems that are not directly faced by the active members and/or the problems that are not severe enough to attract users with an itch will remain unaddressed.
\Dfs, while not a silver bullet to solving the need for telemetry, can act as a middle ground to offer more information in addition to the vocal majority.
The additional information shared by the community can help with discussions to provide a broader but not absolute view of how the tools are used in the community.
This information can guide the process of creating a ``sane default'' setting.
However, even if this information is provided, we may still face a divided community on what constitutes a ``sane default''.
Some previous studies have explored the idea of leveraging community configurations to help with creating better configurations for production software~\cite{zheng2011massconf,talwadker2019popcon}.
We believe that there are unique challenges faced in \dfs.
Unlike software deployed in production environments (e.g., databases, web servers), user-facing software often does not have an optimal configuration.

\opportunity{Leveraging \dfs to help creating advanced configuration recipes.}
Configurations for software tools can get complicated.
For example, it is common to observe \emacs configurations with thousands of lines of \emph{elisp} code.
Developers often learn from other developers and gradually add and improve their own configuration.
We believe that by leveraging the corpus of \dfs, we can help create better documentation to help developers configure their tools.

A similar concept can be observed in many successful open source libraries, where a section called the ``cookbook'' can be found in the documentation.
The section provides recipes for common use-cases, and serves as a starting point for developers who wish to learn how use the library effectively.
However, creating and maintaining the recipes can be a challenging task.
We believe that \dfs can be a valuable source for creating ``cookbook'' recipes for software tools.
By extracting common real-world configurations from \dfs, we can improve the examples in documentation.
For example, by extracting common configurations for editing Python from \vim configurations, we can provide extended real-world examples to help future developers to configure the software.

\section{Related Work}

Most of the existing research in this broad area focuses on studying software configurations either during the build process (e.g., enabling different features for different configurations of software), or on the configuration of the deployment and run-time environment of production software (e.g., database systems, web servers).
The study of \dfs is also closely related to improving the workflow of developers, since \dfs are used to customize the tools used during development which can help improve the efficiency of developers.
In this work, our focus has been on the user-specific configuration files that users keep to configure the software they use.
Below are some related work to this topic focusing on software configurations and on improving developer workflow.

% \mike{I don't understand the next sentence, esp. ``both'' and ``the scene of developer life''.}
% Since they are both configuration files, and more focused on the scene of developer life, in this section, we include some related work in both software configuration, and improving developer workflow.

\subsection{Software Configurations}

%-- While many studies have looked at configurations, the main focus has been on software build configurations and software run-time configurations~\cite{nadi2015configuration,zhou2015extracting,nair2018finding,xu2015hey,xu2013not,zheng2011massconf}.
While many studies have looked at the broad topic of software configurations, the main focus has been on two areas:  build configurations and run-time configurations that model user preferences~\cite{nadi2015configuration,zhou2015extracting,nair2018finding,xu2015hey,xu2013not,zheng2011massconf}.
These studies often focus on production software with the aim to improve the build process or optimize performance.
This includes studying the build configuration files.
Nadi \etal{}~\cite{nadi2015configuration} proposed a static analysis approach to extract configuration constraints from software build files (e.g., \texttt{Makefile}s).
Their methods are able to provide insight from creating a variability model from the constraints and reason about the build configuration.
Zhou \etal{}~\cite{zhou2015extracting} proposed a method to parse build files such as \texttt{Makefile}s and detect potential problems in the configuration space for software using symbolic execution.

Other research focuses on run-time configurations.
Many software systems that are run in production environments are highly customizable to satisfy different needs.
Unlike build-time configurations which cannot be changed after the software is built, run-time configurations are more flexible.
Databases are a class of software that fits the description well, and it is no surprise that they are also one of the most studied systems for run-time configurations~\cite{duan2009tuning}.
One goal of tweaking run-time configurations is to have the software system running at optimal settings.
Nair \etal{}~\cite{nair2018finding} proposed a sequential model-based method that explores the configuration space and tries to determine the next best configuration.
M{\"u}hlbauer \etal{}~\cite{muhlbauer2019accurate} showed that Gaussian process models can accurately estimate the performance-evolution history of real-world software systems.
Kaltenecker \etal{}~\cite{kaltenecker2019distance} showed that from empirical evaluation, distance-based sampling on configuration space can yield more accurate performance models for medium to large sample sets.

Studies have also shown that the design of configuration ``knobs'' are sub-optimal.
Xu \etal{}~\cite{xu2015hey} investigated on the complexity of configuration settings on multiple software systems and shows that only a small percentage of configurations are altered by users while a significant percentage of the configurations are never changed.
Sayagh \etal{}~\cite{sayagh2020software} interviewed and surveyed developers on the design of run-time software configuration options suggesting that software configurations are often added unplanned and unorganized.
Xu \etal{}~\cite{xu2013not} proposed a tool to support users by automatically infer configuration requirements through mining constraints from the configuration space.

% relationship to our study
Unlike previous research which heavily focuses on configurations for production software,
in this study we focus on \dfs from the user space.
Compared to production software, many configurations in the user space do not have a strong focus on performance or have a definitive way of specifying configuration.
User-specific configurations are often tailored to the needs of individual developers and often evolve over time.

% =============================================================================

\subsection{Developer Workflow}

Developers' time and interest is a valuable resource.
Many studies have focused on how to develop or improve tools to increase developer productivity.
More importantly, studies have shown that tool choice does matter for developers~\cite{kavaler2019tool}.
And researchers have also found surprising ways of how developers discover new tools~\cite{murphy2019developers}.

Current research often focuses on one particular tool or a specific set of tools.
For example, Schr{\"o}der \etal{}~\cite{schroder2022empirical} performed an empirical study on aliases used during command-line customization.
Their study suggests that developers mainly use aliases for shortcuts, modifications, and scripting.
Johnson \etal{}~\cite{johnson2013don} investigated the reasons why developers look away from existing static analysis tools hoping to gain insight on how to improve them.
In another study, Damevski \etal~\cite{damevski2016mining} studied developer interactions in Visual Studio to detect potential usage problems.
These efforts have the same goal to improve the developer efficiency, however choosing and adopting the best practices and tools is a challenging task for developers~\cite{snipes2013towards}.
Some attempts have been performed to tackle this issue; for example, in a follow up study, Snipes \etal~\cite{snipes2014experiences} explored the possibility to gamify the process and received dividing feedback from a pre-study survey.

% relationship to our study
In our study, we focused on the \dfs management by developers.
Since \dfs is a collection of user-specific configurations, instead of focusing on a specific aspect and on a specific purpose, our study is a high-level view of how these configuration are managed and maintained over time.

\section{Threats to Validity}

\paragraph{Internal Validity}
When deciding to create the \dotdataset, we set strict filters to the repositories selected.
This suggests that we can miss out on repositories that may be of interest.
However, we selected these criteria to ensure that we are looking at the \dotrepos owned by developers, which we also verified and confirmed the effectiveness of the criterion in RQ1.
During our process to determine the occupation of \dotrepos owners, we leveraged only the owner's public profile as well as a simple user name search; in only a few cases were we unable to determine the occupation to our satisfaction.
It is possible that the owner may as well work in fields that require developing
software.
Since a more thorough search, such as using the owner's commit email, may violate the privacy of the owner, we decided stopping at the stage of viewing the public profiles.

We chose to leverage the \kshape algorithm to extract the \dfs maintenance patterns modeled as a time series based on historical code churn.
\kshape focuses on the shape attributes of the time-series and is an alternative approach compared to extracting features from the code commits.
Using a shape-based algorithm allows us to avoid the downsides and potential problems of determining the important features to extract, and also to have a higher focus on the activity trends of the code change alone.
Since this is still a fairly new approach, in future work, research can investigate the differences and effectiveness of the different approaches for clustering code churn.

\paragraph{External Validity}
We leveraged \emph{GHTorrent} as our basis for selecting and collecting \dotrepos from \github.
However, \emph{GHTorrent} contains data only up to 2019 and has not been actively maintained or updated since then.
While leveraging \github API can replicate part of the functionality, as implied by Jungwirth \etal{}'s work~\cite{jungwirth2023connecting}, it is unable to match the completeness of what \emph{GHTorrent} collected due to the limitation of the API.
While \github is largely considered as the \emph{de facto} location for developers to host their projects, in recent years, many other options have gained popularity.
Some options are also centralized hosting a large variety of projects such as \emph{GitLab}.
Meanwhile, self-hosted options also exist, where an independent \github-like instance can be hosted for a specific developer or organization.
One example would be \emph{Savannah}\footnote{\url{https://savannah.gnu.org/}}, the place where \emph{GNU} software is hosted.
Due to security reasons, developers are also unlikely to host their work-related user-specific configuration files on publicly.
Therefore, our results may be biased towards open source applications and non-professional settings.

Because they originate in the \nix world of the 1970s --- largely before the advent of graphical user interface --- \dfs have historically had a heavy focus on the \emph{CLI} applications.
This can also be observed from the set of common \dfs across repositories.
Since \emph{GUI} applications may have different ways of storing configuration (e.g., a binary file), and have built-in syncing functionality (e.g., \vscode), they may not be captures in the \dotrepos.
Storing configurations in plain-text may be replaced by other means in the future, however, we believe the requirement for customization still exists.
Future research can study configurations for \emph{GUI} software and compare the results with our work.

% \mike{My memory is that sharing settings really took off when people got workstations with big graphical displays and started using X Windows.  The default app settings were often lousy, and people in shared lab spaces would ask ``How did you get it to do that'' WRT settings.   This was before online shared repos became a thing tho.  Sharing was informal or done thru USENET.  In the pre-Internet days, we used to joke that "UNIX is an oral tradition'' because there were basically no written resources for it apart from man pages.}

\section{Summary}

In this work, we study the practice of sharing and maintaining \dfs based on \dotrepos collected from \github.
We observe that sharing \dfs is a common practice among developers, with configuration files for text editors, shells and \git being the most common.
While developers track many \dfs, only a small amount of the \dfs are constantly updated.
We extracted code churn history patterns from frequently updated \dfs and find that there is no significant relationship between the code churn history pattern and the type of \dfs.
We discuss the challenges developers face in managing \dfs and how we can leverage the publicly shared \dfs to help creating ``sane defaults'' and constructing ``cookbook'' recipes for documentation.

\section*{Conflict of Interests}
The authors declared that they have no conflict of interest.

\section*{Data Availability Statement}
A dataset consists of the \dfs analyzed are available on Zenodo.\footnote{\url{https://zenodo.org/record/8368471}}

\bibliographystyle{ieeetr}
\bibliography{references}

\end{document}